\documentclass[preprint,12pt]{elsarticle}
\usepackage{latexsym} 
\usepackage{mathptm}
\usepackage{amsmath}
\usepackage{amssymb}
\def\beq{\begin{equation}}
\def\eeq{\end{equation}}
\def\beqn{\begin{eqnarray}}
\def\eeqn{\end{eqnarray}}
\journal{Physics Letters B}

\begin{document}

\begin{frontmatter}
 
\title{A possibility to solve the problems with quantizing gravity}
\author{Sabine Hossenfelder} 
\address{Nordita, Roslagstullsbacken 23, 106 91 Stockholm, Sweden}
\begin{abstract}
It is generally believed that quantum gravity is necessary to resolve the known tensions
between general relativity and the quantum field theories of the standard model. Since 
perturbatively quantized gravity is non-renormalizable, the problem how to unify all interactions
in a common framework has been open since the 1930s. Here, I propose a possibility to circumvent 
the known problems with quantizing
gravity, as well as the known problems with leaving it unquantized: By changing the
prescription for second quantization, a perturbative quantization of gravity is 
sufficient as an effective theory because matter becomes classical
before the perturbative expansion breaks down. This is achieved by considering
the vanishing commutator between a field and its conjugated momentum as a symmetry
that is broken at low temperatures, and by this generates the quantum phase that
we currently live in, while at high temperatures Planck's constant goes to zero.
\end{abstract}
\begin{keyword}
Quantum gravity, Semi-classical gravity

\end{keyword}

\end{frontmatter}

\section{Why Quantize Gravity?}

The gravitational interaction stands apart from the other interactions of the standard model by its
refusal to be quantized. This still missing theory of quantum gravity is believed necessary to complete
our understanding of nature. Strictly speaking, quantizing gravity is not the problem -- gravity can
be perturbatively quantized. The problem is that the so quantized theory is perturbatively non-renormalizable
and cannot be understood as a fundamental theory. It breaks down at high energies when quantum
gravity would be most interesting.

The attempt to find a theory of quantum gravity has lead to many proposals, but progress 
has been slow. Absent experimental evidence, reasons for the necessity of quantum gravity are
theoretical, most notably:

\begin{enumerate}
\item Classical general relativity
predicts the formation of singularities, infinite energy densities, under quite general circumstances. 
Such singularities are unphysical and should not occur in a fundamentally meaningful theory. 

\item Quantum
field theory in a curved background leads to black hole evaporation. Black hole evaporation
however seems to violate unitary which is incompatible with quantum mechanics. 
It is widely believed that quantum gravitational effects restore unitarity and
information is conserved. 

\item There is no known consistent way to couple a quantum
field to a classical field, and since all quantum fields carry energy they all need to
couple to the gravitational field. As Hannah and Eppley have argued \cite{EH}, the attempt to make such a 
coupling leads either to a violation of the uncertainty principle (and thus would necessitate a 
change of the quantum theory) or to the possibility of superluminal signaling, which brings more
problems than it solves. Mattingly has argued \cite{Mattingly} that Hannah and Eppley's thought experiment
can't be carried out in our universe, but that doesn't solve the problem of consistency.
\end{enumerate}

These points have all been extensively studied and discussed in the literature. The
most obvious way out seems to be a non-perturbative theory, and several attempts to construct one are under way. 

It is worthwhile for the following to identify the problems with coupling a classical to
a quantum field. 

The one problem, as illuminated by Hannah and Eppley is that the fields
would have different uncertainty relations, and their coupling would require
a modification of the quantum theory. Just coupling them as they are leads to an inconsistent
theory. The beauty of Hannah and Eppley's tought argument is its generality, but that
is also its shortcoming, because it does not tell us how a suitable modification of
quantum theory could allow such a coupling to be consistent.

The second problem is that it is unclear how, mathematically, the coupling should be realized,
as the quantum field is operator-valued and the classical field is a function on space-time. 
One possible answer to this is that any function can be identified with an operator on
the Hilbert space by multiplying it with the identity. However, the associated operators 
would always be commuting, so they are of limited use to construct a geometrical quantity
that can be set equal to the operator of the stress-energy-tensor ({\sc SET}) of the quantum fields.

Another way to realize the coupling is to extract a classical field from the
operator of the {\sc SET} by taking the expecation value. The problem
with this approach is that the expectation value may differ before and after measurement,
which then conflicts with local conservation laws in general relativity. Coupling the
classical field to the {\sc SET}'s expectation value is thus usually
considered valid only in the approximation when superpositions carry negligible amounts of
energy.

These difficulties can be circumvented by changing the quantization condition
in such a way that gravity can be perturbatively quantized at low energies, but at 
energies above the Planck energy -- energies so high that the perturbative expansion would
break down -- it becomes classical and decouples from the matter fields. The mechanism
for this is making Planck's constant into a field that undergoes symmetry breaking and induces
a transition from classical to quantum. 
In three dimensions, Newton's constant is  $G= \hbar c/m_{\rm Pl}^2$, so if we keep mass units fix, 
$G$ will go to zero 
together with $\hbar$, thus decoupling gravity. 

It should be emphasized that the ansatz
proposed here does not
renormalize perturbatively quantized gravity, but rather replaces it with a different
theory that however reproduces the perturbative quantization at low energies by construction. 
We will show in the following how this change of the quantization condition addresses the
above listed three problems.

In the approach considered here, Planck's constant is treated as a field. Since 
the normal, fixed, value of Planck's constant (denoted $\hbar_0$) appears as a
vacuum expectation value, we could divide the field by this vacuum expectation value
to obtain a dimensionless quantity. We will not do this in the following, because it would
make the interpretation less intuitive. It should be pointed out though that the approach
can be reformulated in terms of a dimensionless field because we do have access
to a dimensionful constant that serves as reference. 

In the following we set $c=1$ but 
keep $\hbar$ and Boltzmann's constant $k_{\rm B}$. 
The signature of the metric is $(+,-,-,-)$.

\section{Quantization by Spontaneous Symmetry Breaking}

Consider a massless real scalar field $\phi(x,t)$ with canonically conjugated momentum $\pi_\phi(x,t)$. 
Second quantization can be expressed through the equal time 
canonical commutation relations
\beqn
[ \phi(x,t), \pi_\phi(y,t)] = {\rm i} \hbar \delta^3(x-y) \quad, \label{etcr}
\eeqn
or, equivalently, by the commutation relations for annihilation and creation operators. 
If $\hbar = 0$, they commute and one is dealing with a classical field. 

The dimension of a scalar field is most easily found by noting that the kinetic 
term $\partial_\nu \phi \partial^\nu \phi$ in the Lagrangian
should have dimension of an energy density, so that the integral over space-time has the dimension of an action. 
One has a freedom here whether a constant is in front of this term. If one is dealing
with a quantum theory, one often puts an $\hbar^2$ there because then each derivative together with an
$\hbar$ gives a momentum. Since we eventually want to make contact to a classical theory, we will
not put any $\hbar$'s in front of the kinetic term, but instead chose the classical convention from the start on.

In three spatial dimensions this mean then that the dimension of the Lagrangian is $[E/L^3]$, where $E$ denotes energy and
$L$ denotes length. So the dimension 
of $\phi$ is $[E^{1/2} L^{-1/2}]$, and that of the conjugated momentum $\pi_\phi$ is $[E^{1/2} L^{-3/2}]$. 
The Lagrangian is 
\beqn
{\cal L} = \frac{1}{2} \partial^\nu \phi \partial_\nu \phi \quad, \label{lag}
\eeqn
without additional factors, and $\pi_\phi = \partial {\cal L}/\partial \dot \phi$ as usual. 
With this convention, what is in the quantum theory referred 
to as the `mass term' has actually 
the form $\phi^2/l_*^2$, where $l_*$ is a length scale. This is because the unquantized field is not a priori 
associated with particles that could be assigned a mass.

With that dimensional consideration, let us now look at the quantization prescription from a new 
perspective. The requirement that the fields commute, $\phi(x)\pi(y) = \pi(y) \phi(x)$, can be
understood as a symmetry, where we find the classical theory as the symmetric, commuting, phase and 
quantum mechanics in the phase with broken symmetry where the fields do not commute.

In the familiar language of spontaneous symmetry breaking we parameterize
\beqn
[ \phi(x,t), \pi_\phi(y,t)] = {\rm i} \frac{\hbar^2_0}{m_*} \alpha(x,t) \delta^3(x-y) \quad. \label{etcra}
\eeqn
Here, $\hbar_0$ is the normal (measured) value of Planck's constant, $m_*$ is a constant of dimension mass, and $\alpha(x,t)$ is a 
real scalar field. With a Fourier-transform, one can express this in terms of annihilation
and creation operators for $\phi$ of the form 
\beqn
[a_{\vec k}, a^\dag_{{\vec k}'}] = {\rm i} \frac{\hbar_0^2}{m_*} 
\left( \frac{\omega_{k'} + \omega_{k}}{2 \sqrt{\omega_{k'} \omega_k}} \right) \tilde \alpha(\vec k- \vec k')
\eeqn 
where $(\omega_k,\vec k)$ is the wavevector, $\omega^2_k = |\vec k|^2$, and $\tilde \alpha$ is the
Fourier transform of $\alpha$
\beqn
\tilde \alpha(\vec k) = \int d^3 x~ \alpha(x,t) \exp (i x \cdot k) \quad. \label{aadag}
\eeqn 
If $\alpha$ is constant, then $\tilde \alpha$ reduces to the usual delta-function. For the operations
leading to the above expressions, we have merely used the definition of scalar products that are
not affected by the modification of the quantization condition.

We have distributed the dimensionful constant so that $\alpha$ has no prefactor in the path
integral, ie the dimension of $\alpha$ is $[1/L]$. We have further chosen to keep a mass scale fixed rather than a 
length scale for the reason indicated in the introduction. We can then
write $\hbar(x,t) = \hbar^2_0 \alpha/ m_*$ for Planck's constant, which is now a field. The field
$\alpha$ itself is quantized according to the same prescription as $\phi$ up to constants, ie 
$[\alpha(x,t), \pi_\alpha(y,t)] = i \hbar_0 \alpha(x,t) \delta^3(x-y)$. If we expand $\alpha$ in
annihilation and creation operators, these too obey the commutation relation (\ref{aadag}) with
a different prefactor. 

Now we add a kinetic term for $\alpha$ and a symmetry breaking potential, so that the transition amplitude
in the path integral is $\exp( - {\rm i} \hat S)$, where the hat indicates a dimensionless quantity, and
\beqn
{\hat S} &=& \int {\rm d}^4 x \sqrt{-g} \left( \frac{m_*}{2 \alpha \hbar_0^2} \partial^\nu \phi \partial_\nu \phi + 
\frac{1}{2} \partial^\nu \alpha \partial_\nu \alpha -   V(\alpha)/\hbar_0 \right) \nonumber \\
V(\alpha) &=& - 2 \frac{m^2_*}{\hbar_0} \alpha^2(x) \label{normact} 
+ \hbar_0 \alpha^4(x) \quad.
\eeqn
The minima of the potential are at $\alpha =  \pm m_*/h_0$, and 
we live in the vacuum with $\hbar = \hbar_0$. (In order to render the
kinetic term of $\phi$ symmetric under a sign change of $\alpha$, one
would have to also change the sign of $m_*$ along with $\alpha$.
This makes the example of the scalar field somewhat unappealing, 
but we will see below that this is not necessary for the case of
gravity, which is what we are actually interested in.)

As mentioned in the introduction, to obtain a dimensionless
quantity whose value is meaningful regardless of units, one could now normalize
the field $\hbar$ to $\hbar_0$. However, this would make the interpretation
of the appearing quantities less intuitive, so we will not do this redefinition here.
The previously made statement that the failure of the field and its momentum
to commute represents a breaking of symmetry should be understood as a rephrasing 
of the breaking of symmetry in the ground state of the above potential.

This whole exercise can be summarized by saying that we've chosen the
dimensions of the fields so that in the path integral the field $\alpha$, which determines
Planck's constant, appears in the usual form.

Thus, we are left with an action of two scalar fields, where the symmetry breaking
part works as normally. The only unusual is the quantization condition for the the fields. Symmetry breaking 
then happens with a drop of temperature because the minima of the potential
change with the temperature \cite{Kapusta}. In the limit $T \gg T_c =2 m_* k_{\rm B}$, the finite temperature corrected potential 
receives additional terms
\beqn
V(T,\alpha) = V(\alpha) + \frac{1}{2\hbar_0} (k_{\rm B} T)^2 \alpha^2 - \frac{\pi^2}{90} n k_{\rm B} T + \dots\quad, \label{ftv}
\eeqn 
where $k_{\rm B}$ is Boltzmann's constant, $n$ is the number-density of the field, and the dots indicate terms of higher 
order in $\hbar_0$. Note that they are indeed higher order in $\hbar_0$ and not in $\hbar$ because there's no additional
$\hbar$ in the potential. The first correction term can be interpreted as a temperature-dependent
mass term for the $\alpha$-field that counteracts the negative mass term in the potential. The second correction term
is the free energy of a massless spin-0 boson. 

To understand what happens if the symmetry is broken, we first note that in principle the temperature dependence
of the potential is not a quantum effect, it is an in-medium effect. However, Planck's constant appears in statistical
mechanics as the normalization constant for the measure of momentum space. Taking it to zero does not create a classical
limit, but instead ill-defined quantities. One frequently considers limits in which Planck's constant is small compared
to some other quantity of dimension action. That is formally similar to changing Planck's constant, but has a very
different physical meaning. We don't want to look at the limit where Planck's constant is small compared to some
other value, we want to change it. This is not a process that statistical thermodynamics in its standard form deals
with.

Eg the free energy of the bosonic gas that appears in the potential is normally written 
$(kT)^4/\hbar^3$, which seems to diverge with $\hbar$ to zero. But to obtain a meaningful limit, we have to
keep a physically meaningful quantity fixed. As the above notation already suggests, we opt to keep the number density fixed 
when varying $\alpha$ but not any other variable, ie
\beqn
\frac{\partial n}{\partial \alpha} \Big|_{(T,S,p)} = 0 \quad. \label{nfix}
\eeqn
This does of course not mean the number density is constant. To begin with it depends on
the temperature. What Eq (\ref{nfix}) expresses is that there is no additional temperature-dependence in $n$ from the
variation of $\alpha$. In other words, we identify $n = (kT/\hbar_0)^3$. 

There is another way to look at this limit. Planck's constant is the measure in phase space, and  can be 
understood as a product of a length for
the coordinate space and an energy for momentum space. We have introduced a mass scale in the beginning,
which we keep fix, so then the unit for the length has to go to zero with Planck's constant, which
is why the number-density would diverge in these units. This unphysical
behavior is cured by choosing a constant unit of length instead, which has the same effect 
as putting $\hbar_0$ in the denominator of $n$. 

The potential for the quantized scalar field further receives corrections from quantum fluctuations which leads to the 
Coleman-Weinberg effective potential. In the usual case for a $\phi^4$-interaction, the loop
corrections become large for small $\phi$ and therefore quantum fluctuations can break the classical symmetry. 
Since in our case the potential for the field $\alpha$ takes the usual form, the same can happen here.  It is
less clear what happens with the loop corrections in the high-temperature limit when the modification
of the quantization condition becomes relevant.

As mentioned above, what we actually want to unquantize is not a $\phi^4$ theory, but gravity. In this
case, the normalized action is
\beqn
\hat S = \int {\rm d}^4 x \sqrt{-g} \left( \frac{m_{\rm Pl}^2 m_*^3}{\alpha^2 \hbar_0^4} {\cal R} + \frac{1}{2} \partial^\nu \alpha \partial_\nu \alpha -   V(\alpha)/\hbar_0  \right) ~.
\eeqn
Other matter fields can be added like the scalar field example. 
This looks very much like Brans-Dicke, but has a different
kinetic term and the symmetry breaking potential. The most relevant
difference though is the quantization condition. (Note that the first
term does not change sign under $\alpha \to - \alpha$.)

In summary, for this to work, one needs the normalized action of the form (\ref{normact}) together with the quantization
postulate (\ref{etcra}). Since the quantization postulate is essentially a constraint on the fields, one could
alternatively add it with a Lagrange multiplier to the Lagrangian. That might not be particularly elegant, but the point
is here just to show that it can be done. 

\section{How unquantization addresses the problems with quantizing gravity}

The
coupling constant of gravity to matter in the perturbative quantization is $\sqrt{G}=\sqrt{\hbar}/m_{\rm Pl}$
and thus goes to zero with $\hbar \to 0$, provided the Planck mass is held fix. This limit was previously discussed from a different
perspective in \cite{KowalskiGlikman:2006vx,Smolin:2010xa} as being of
interest as a non-quantum relic of quantum gravity in the limit $\hbar, G \to 0$. Here, we suggested to not only consider this a
corner of the parameter space, but a limit that is actually realized at high energies. 

\begin{figure}[ht]
\includegraphics[width=5.5cm]{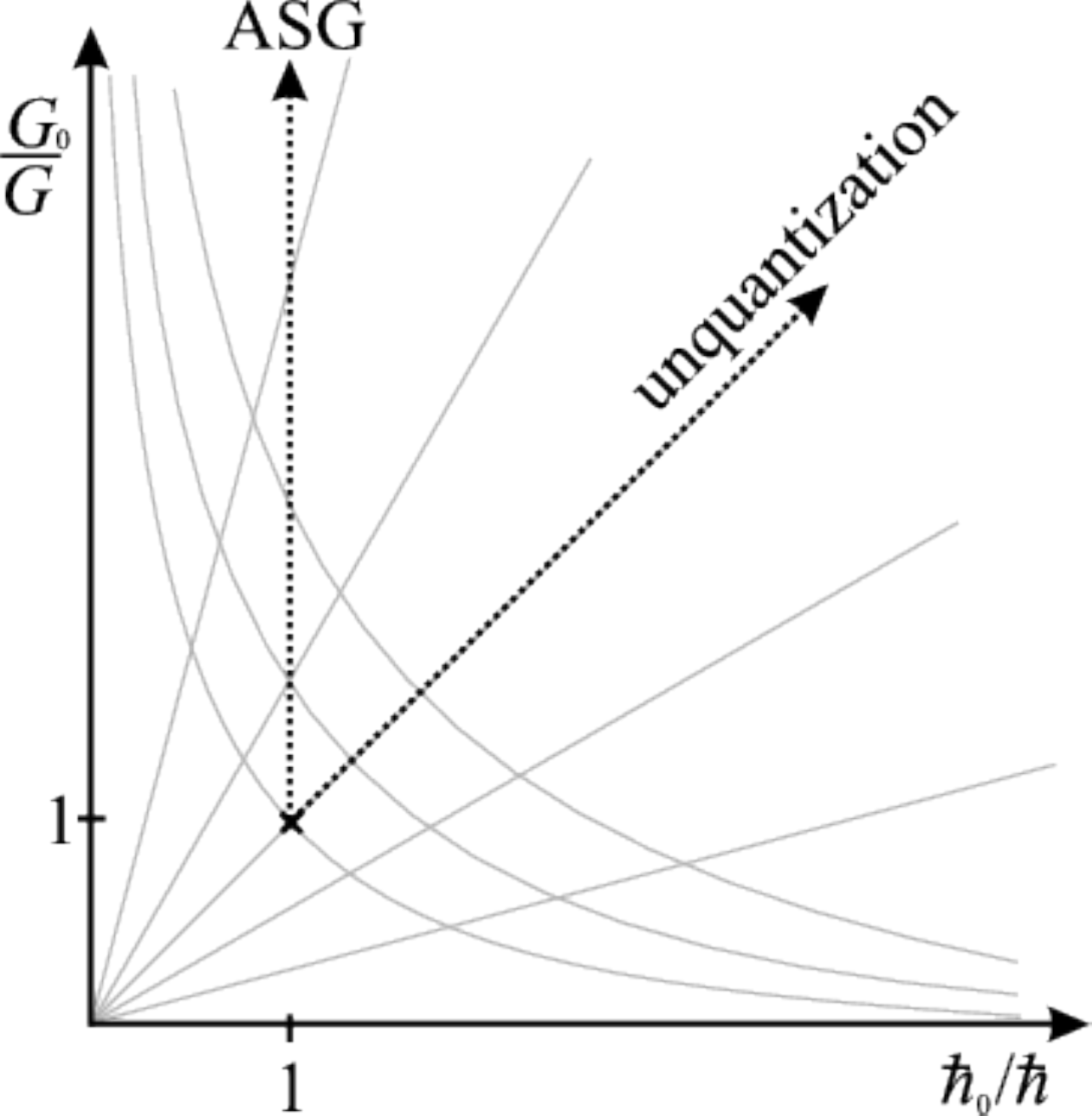}
\caption{{\small The parameter space for Newton's constant $G$ and Planck's constant $\hbar$, normalized
to the measured values $G_0$ and $\hbar_0$. The black cross indicates the familiar low-energy theory, 
the place where quantization of general relativity is perturbatively non-renormalizable. 
Shown in grey are the hyperbolas where the Planck length is constant, and the straight lines where
Planck's mass is constant. In Asymptotically Safe Gravity (ASG), Newton's constant decreases at 
high energies, weakening gravity. In the here discussed scenario of unquantization, Planck's
constant goes also to zero. }}
\label{1}
\end{figure}

The weakening of gravity at high energies can also be found in scenarios where gravity
is asymptotically safe. The difference between both cases is most easily shown by a figure,
see Fig \ref{1}. 

Let us now revisit the motivations for the need for quantum gravity:

\begin{enumerate}
\item The formation of singularities: If matter is compressed, it eventually forms a degenerate Fermi
gas. If it collapses to a black hole, it collapses rapidly and after horizon
formation lightcones topple inward, so no heat exchange with the environment can take place and
the process is adiabatic. The entropy of the degenerate Fermi gas is proportional to $T n^{-2/3}$,
which means that if the number density rises and entropy remains constant, the temperature has to rise 
\cite{Kothari}. If the temperature rises, then gravity eventually decouples and there is 
no force left to drive the formation of singularities. Towards the Big Bang singularity, temperature
also raises and the same conclusion applies. Note that the unquantizing field makes a 
contribution to the source term, necessary for energy conservation.

\item The black hole information loss: If there is no singularity, there is no
information loss problem \cite{Hossenfelder:2009xq}. Moreover, if the matter inside the apparent horizon 
becomes classical and has no Pauli 
exclusion principle, there's
nothing preventing a remnant from storing large amounts of information that can be very suddenly
released once the black hole has evaporated enough. 

\item The problem of coupling a classical
to a quantum theory: There is never a classical field
coupled to a quantized field. There is a phase when both are quantized and coupled, and one
when both are unquantized and decoupled. One might say that fundamentally the fields are neither classical
nor quantum in the same sense that water is fundamentally neither liquid nor solid.
\end{enumerate}

It remains to be addressed what happens to the divergence of the perturbative expansion of 
perturbatively quantized gravity. This is necessary to understand what happens in a highly
energetic scattering event where the mean field approximation that is made use of for the
symmetry restoration with temperature is not applicable. 

We first note that at high energies the operator that contributes the negative mass term to
the potential of $\alpha$ becomes less relevant than the quartic term, so at least asymptotically
the symmetry should be restored in the sense that the vacuum expectation value of $\hbar$ goes
to zero. With the field rescaling that we have used here (see also \cite{Brodsky:2010zk}), the perturbative 
expansion is also an expansion in $\hbar$,  but it is not quite as easy as this for the following reason.
 
Consider an $S$-matrix expansion within the above model. The expansion of the $S$-matrix works
as usual, but the modification comes into play when we look at a transition amplitude of that
 $S$-matrix with some interaction vertices. It is evaluated by
using the commutation relations repeatedly until annihilation operators are shifted to the very
right side, acting on the vacuum, which leaves $c$-numbers (or the Feynman rules respectively). Now,
if Planck's constant is a field,
every time we use the commutation relation, we get
a power of $\alpha$ and the respective factors of the constant $h_0$ and $m_*$. So, in the end we 
do not have powers of the vacuum expectation value of $\tilde \alpha$, but the expectation value of powers 
of $\tilde \alpha$. We thus need to take a second step,
that is using the commutation relations on $\tilde \alpha$ itself. But exchanging any two terms in
the expansion of $\tilde \alpha$ will only generate one new $\tilde \alpha$ from the commutator (and the
respective powers of $\hbar_0$). One
can thus get rid of the expectation value of powers, so that in the end we will have a
series in $h_0$ and vacuum expectation values of $\tilde \alpha$. 

If $\alpha$ goes to zero only for infinitely large energies, then one cannot tell if the
series is finite without further investigation of its convergence properties. However, if we
consider the symmetry breaking potential to be induced by quantum corrections at low order,
the transition to full symmetry restoration may be at a finite value of energy. In this case
then, the quantum corrections which would normally diverge would cleanly go to zero, removing
this last problem with the perturbative quantization of gravity.

We have discussed here a model based on a modification of the quantization
condition that we have interpreted as a procedure of unquantization. This
modification should however primarily be understood as a motivation for the 
model, and as a guide for the interpretation of its effects. It is possible that 
the approach can be reformulated in other ways that offer a different 
interpretation of the behavior.

The solution proposed here has the potential to address a long standing problem in theoretical
physics. To be successful in that however, a closer investigation is required. There is, most
importantly, the
question of experimental contraints from coupling the scalar field that is Planck's constant
to gravity, which might lead to a modification of general relativity and observable
consequences. It also remains to be seen if a concrete example for the symmetry breaking
can be constructed in which it can be shown explicity, and beyond the general
argument for such a possibility given above, that the perturbation series converges. 
The beta-function of the model and its relation to the case of
Asymptotically Safe Gravity is of key interest here. And while the avoidance of the
Big Bang and black hole singularities are the most relevant cases for our universe, it remains
to be seen if a more widely applicable statement can be derived that addresses the singularity
theorems in general.

\section*{Acknowledgements}

I thank Cole Miller, Roberto Percacci and Stefan Scherer for helpful discussions.

\end{document}